\documentclass{iau}
\usepackage{graphicx}
\usepackage{hyperref}

\emergencystretch=\maxdimen
\title[About the A\&M databases in the planetary community] 
{About the atomic and molecular databases in the planetary community -- A contribution in the Laboratory Astrophysics Data WG IAU 2022 GA session}

\author[M. Rengel]   
{M.~Rengel$^1$}

\affiliation{$^1$Max-Planck-Institut f{\"u}r Sonnensystem Forschung, Justus-von-Liebig-Weg 3, 37077 G{\"o}ttingen, Germany, \\ email: {\tt rengel@mps.mpg.de}
}

\pubyear{2022}
\volume{371}  
\pagerange{119--126}
\jname{IAU 371 Symposium}
\editors{D. Soderblom \& G. Nave, eds.}

\begin{document}

\maketitle

\begin{abstract}
This paper corresponds to an invited oral contribution to the session 5A organised by the IAU inter-commission B2-B5 working group (WG) “Laboratory Astrophysics Data Compilation, Validation and Standardization: from the Laboratory to FAIR usage in the Astronomical Community” at the IAU 2022 General Assembly (GA) \cite{talk}. This WG provides a platform where to discuss the Findability, Accessibility,Interoperability, Reuse (FAIR) usage of laboratory Atomic and Molecular (A\&M) data in astronomy and astrophysics.

A\&M data play a key role in the understanding of the physics and chemistry of processes in several research topics, including planetary science and interdisciplinary research in particular the atmospheres of planets and planetary explorations, etc. Databases, compilation of spectroscopic parameters, and facility tools are used by computer codes to interpret spectroscopic observations and simulate them. In this talk I presented existing A\&M databases of interest to the planetary community focusing on access, organisation, infrastructures, limitations and issues, etc.


\end{abstract}


\section{Introduction}
The talk represented a tour on A\&M databases used in the planetary and exoplanetary communities not from the perspective of a developer, but of an user.  
A\&M databases compile and provide detailed spectral information for A\&M to feed codes that predict and simulate radiation in gaseous media.
 Between the applications, we find atmospheric physics: (exo) planetary atmospheres, comets and small bodies. These databases are a critical input for the codes which predict and interpret spectra of planetary atmospheres (hydrostatic equilibrium atmospheres and expanding comas), and space and ground-based telescopes facilities depend on the quality and extent of reference A\&M parameters.
 
Line lists typically contain hundreds to billions of individual transitions. A\&M databases contains detailed A\&M spectroscopic parameters of A\&M like pressure-broadening (shapes), collision-induced absorption (CIA), transition intensity or cross sections, line  shape  parameters, rotation-vibration transition position (wavelength, frequency), parameters  to  describe  how  these  vary  with  temperature  and  pressure, aerosol   indices   of   refraction, microphysical and optical properties   of  atmospheric aerosols, for example. 

\section{Some existing A\&M databases - Data and file formats}
Several groups worldwide generate and compile A\&M data through measurement and/or calculation (e.g. {HITRAN}\footnote{\href{https://hitran.org}{https://hitran.org}}[\cite{Gordon22}], {GEISA}\footnote{\href{https://geisa.aeris-data.fr/}{https://geisa.aeris-data.fr/}}[\cite{jacquinet2016}], {JPL Molecular Spectroscopy}\footnote{\href{https://spec.jpl.nasa.gov}{https://spec.jpl.nasa.gov}}, {CDMS}\footnote{\href{https://cdms.astro.uni-koeln.de}{https://cdms.astro.uni-koeln.de}}[\cite{endres2016}], {VAMDC} \footnote{\href{https://vamdc.org}{https://vamdc.org}}[\cite{dubernet2010virtual, dubernet2016Virtual, decadeWithVAMDC}], 
{ExoMol}\footnote{\href{https://www.exomol.com}{https://www.exomol.com}}[\cite{Tennyson20}], {HITEMP}\footnote{\href{https://hitran.org/hitemp}{https://hitran.org/hitemp}}[\cite{rothman2010}], 
{VALD} \href{http://vald.astro.uu.se}{http//vald.astro.uu.se}, 
{MoLLIST} \href{http://bernath.uwaterloo.ca/molecularlists.php}{http://bernath.uwaterloo.ca/molecularlists.php}, {Ames Molecular Spectroscopic Data for Astrophysical and Atmospheric Studies} (\href{http://huang.seti.org}{http://huang.seti.org}, 
{TheoReTs} \href{https://theorets.univ-reims.fr/}{https://theorets.univ-reims.fr/}, etc.).  Several secondary databases and information services are fed with data from such sources in a fragmented manner. A variety of data formats (cross sections, K-tables, line-by-line, super-lines) and file formats (e.g. .hdf5, .pickle, .mp4, .txt, .npy) are generated. There are online tools such as {HAPI} to extend functionalitties \href{https://hitran.org/hapi/}{https://hitran.org/hapi/}[\cite{kochanov2016}], and {exo-k} library to handle radiative opacities \href{https://pypi.org/project/exo-k/}{https://pypi.org/project/exo-k/}[\cite{leconte21}], that enable conversion between different formats.
As part of the spectroscopic input to atmospheric codes, the HITRAN molecular spectroscopic database is already internationally recognised as standard in the planetary community, and the ExoMol database, valid over extended temperature ranges, is widely used in the exoplanetary community.

\section{Some (exo)planet atmospheric radiative transfer and inversion codes}
One commonly used way to interpret the measured spectra is calculating a synthetic spectrum for comparison with that measured by solving the radiative transfer (forward model) -i.e., computation of the outgoing radiation from the planetary surface for a given set of free parameters-, and inferring parameters like temperature and abundance. This last step is called inversion and consists in 
comparing the measured and best modelled spectra adjusting the atmospheric parameters in such a way as to minimise any discrepancy. A number of radiative transfer codes or forward models and inversion algorithms (retrieval technique) are already generally available and used by the planetary and exoplanetary characterisation communities (Table \ref{table:1}). When calculating a synthetic atmospheric spectrum, lines are read from A\&M databases. The retrieval or parameter fitting techniques commonly used are Optimal Estimation algorithm, nested sampling, Markov chain Monte Carlo method, and Grid search. Some examples of applications of such algorithms are given in [\cite{rengel2008, hartogh2010,rengel2014,shulyak2019,shulyak2020,rengel2022,villanueva2022}]

\tiny
\begin{table}
\caption{Some radiative transfer and inversion codes used in the (exo) planetary community}              
\label{table:1}      
\centering                                      
\begin{tabular}{c c c }          
\hline\hline                        
Code name & Link & Reference \\    
\hline                                   
    ARCiS & \href{http://www.exoclouds.com}{http://www.exoclouds.com} & \cite{min2020}  \\      
    TauREx & \href{https://taurex3-public.readthedocs.io/en/latest/}{https://taurex3-public.readthedocs.io/en/latest/} & \cite{al2021}       \\
    NEMESIS & \href{https://users.ox.ac.uk/~atmp0035/nemesis.html}{https://users.ox.ac.uk/~atmp0035/nemesis.html} & \cite{irwin2008}     \\
    petitRADTRANS & \href{https://petitradtrans.readthedocs.io/en/latest/}{https://petitradtrans.readthedocs.io/en/latest/} & \cite{molliere2019}     \\
    PSG & \href{https://psg.gsfc.nasa.gov/}{https://psg.gsfc.nasa.gov/} & \cite{villanueva}       \\
    CHIMERA & \href{https://github.com/mrline/CHIMERA}{https://github.com/mrline/CHIMERA} & \cite{line}      \\
    PLATON & \href{https://github.com/ideasrule/platon}{https://github.com/ideasrule/platon} & \cite{platon}       \\
    ATMO & \href{https://www.erc-atmo.eu}{https://www.erc-atmo.eu} & \cite{atmo}       \\
    MOLIERE & \cite{moliere2}  & \cite{moliere}       \\
    ARTS & \href{https://radiativetransfer.org/}{https://radiativetransfer.org/} & \cite{arts}       \\
    Home made (MPS) & - & \cite{jarchow}       \\
    Helios-r2 & \href{https://github.com/exoclime/Helios-r2}{https://github.com/exoclime/Helios-r2} & \cite{helios}       \\ 
    SCARLET & - & \cite{benneke}       \\
    BART & \href{https://github.com/exosports/BART}{https://github.com/exosports/BART} & \cite{blecic}       \\
    INARA & \href{https://gitlab.com/frontierdevelopmentlab/astrobiology/inara}{https://gitlab.com/frontierdevelopmentlab/astrobiology/inara} & \cite{inara}       \\
\hline                                             
\end{tabular}
\end{table}

 \normalsize
\section{Discussion: Needs and wish-list }
In spite of the tremendous advances and current efforts in the generation of databases, there are still improvements in progress. A growing demand for spectroscopic data for (exo) planetary studies and other atmospheres is being driven by scientists who are interested in modelling as well as observing diverse bodies.
Line lists are generated from experiments and/or ab initio calculations and may be incomplete or contain errors.  Databases differ in completeness, and some ones do not accurately characterise high-frequency spectral regions. Sometimes there are no datasets for a specific problem at hand.
Atmospheric codes used by the planetary and exoplanetary characterisation communities, that are designed to solve the radiative transfer equation by looking at the propagation of radiation through a medium and simulate observations and infer parameters, have their own methods for the computation of opacities and there are no community standards. Furthermore, there are also no community standards in the selection of atmospheric codes in mission planning. There are needs to increase accessibility of opacities (computation, access, visualisation, manipulation), laboratory measurements of molecular cross-sections, and pressure broadening description for some species, among many other aspects.

Furthermore, going into details, the community identifies needs in the following aspects: isotopologues: CIA (more lists of N${_2}$O, CH${_4}$, SO), expansion of CIA of secondary species for wider temperature and wavelength ranges, additional experiments for CO intensities, improvement of the CH${_4}$ quality of the line shape parameters, continuum absorption by water vapour, partition functions at higher temperatures for all species (around 5000 K), more friendly ways implementing new data in the RT codes, more aerosol refractive indexes above 500-600 K, more saturation vapour pressure data, kinetic data at high temperature, Rayleigh scattering for non-H${_2}$, collisional xsecs (in particular: mid-sized organics, H${_2}$O${^+}$, diatom-H${_2}$, CH${_3}$OH, HCN, SO${_2}$, CH${_2}$) and high-resolution  xsecs (R=1E6+) atmospheres, organic sulphide gases for the line lists (biomarkers).

\acknowledgments{I thank Sergey Yurchenko and Iouli Gordon for the input and discussions. I thank also the members of the SOC of the session, and the members of the inter-commission B2-B5 working group in special to Marie-Lise Dubernet. I acknowledge the support by the DFG priority program SPP 1992 “Exploring the Diversity of Extrasolar Planets” (DFG PR 36 24602/41).

For the purpose of Open Access, a CC-BY-SA 4.0 public copyright licence has been applied by the author to the present document and will be applied to all subsequent versions up to the Author Accepted Manuscript arising from this submission.
}

\end{document}